\documentclass{article}
\newcommand{\bfr}{\begin{flushright}}
\newcommand{\efr}{\end{flushright}}
 
\begin{document}
\title{Hosotani model in closed-string theory
}
\author{Kiyoshi Shiraishi\\
Institute for Nuclear Study, University of Tokyo, Japan
}
\date{Classical and Quantum Gravity {\bf 7} (1990) 135--148
}
\maketitle
\begin{abstract}
The Hosotani mechanism in the closed-string theory with current
algebra symmetry is described by the (old covariant) operator method.
We compare the gauge symmetry breaking mechanism in a string
theory which contains $SU(2)$ symmetry originated from current algebra
with the one in an equivalent compactified closed-string theory. We also
investigate the difference between the Hosotani mechanism and the
Higgs mechanism in closed-string theories by calculation of a
four-point amplitude of `Higgs' bosons at tree level. 
\end{abstract}

\section{Introduction}

String theories  \cite{1} have made a great impact on the development of
unification in forces and matters in nature.
The original models of string theories as candidates for the
unification were formulated in higher dimensional spacetime
\cite{2,3}. In such models, we must consider that the extra dimensions
except for four dimensions become invisible to us \cite{4}.

Compactification of extra spaces is utilised in such theories
to generate gauge symmetries which may govern the forces in
nature \cite{5}. This scheme is the extension of the idea of Kaluza and
Klein \cite{6}. The gauge interaction can be unified with
gravity in higher dimensions. A lot of work has been done on the
generalised Kaluza-Klein theory, including cosmological considerations
\cite{7}.

Recently, it has been clarified that large gauge
symmetries are required in order to cancel quantum anomalies \cite{3}
in string theories. Thus, if we want to construct the unified theory
from string theory in higher dimensions, we should consider
symmetry-breaking mechanisms rather than symmetry generation at
the stage of the compactification into four dimensions. In fact, the
compactification onto the manifolds
\cite{8} (or orbifolds
\cite{9}) which have no, or few, continuous symmetries is considered in
such string theories.

More recently, the investigation by fermionic construction
of string model building \cite{10} and other methods to
formulate new string models \cite{11} has resulted in the concept of
four-dimensional string theory. Most
recently, Gepner
\cite{12} showed that some of the four-dimensional models can be
interpreted as compactified theories on Calabi-Yau manifolds, which
give rise to phenomenologically favoured low-energy aspects \cite{13}.
Thus it could be said that the geometrical
interpretation of compactification `strikes back' in string
theories.

Now let us go back to the issue of gauge symmetry breaking.
The mechanisms for geometrical symmetry breaking are known in field
theory with extra dimensions. The so-called Wilson loop mechanism or
Hosotani mechanism
\cite{14,15} is one of them; the breakdown of symmetry is caused by,
roughly speaking, the `vacuum expectation value' of gauge fields on an
extra, non-simply-connected space. These vacuum gauge fields play the
role of the order parameter in the mechanism. The simplest example for a
non-simply-connected manifold is a circle, $S^1$.  The model with
extra $S^1$ is originally investigated by Hosotani \cite{15} and,
because
of its simplicity, many authors persued the Hosotani
model from various points of view \cite{16,17}.  Of course the
simplest model is not realistic, but it is expected that we can
examine qualitative properties of the mechanism by the study of the
model with the extra space $S^1$. 

Particle physicists often regard the test of their models to be their
explanation of early evolution in the universe. From a cosmological
viewpoint, phase transition in the early universe is an interesting 
subject to investigate (see, for example
\cite{18}). The present author studied the one-loop free energy of the
Hosotani model with several sorts of matter fields at zero and finite
temperature. We concluded that no remarkable dependence on
temperature was found in the model
\cite{17}. Similar analysis was made for a 
model in open-string theory and almost the same
conclusion was drawn \cite{19}. In the case of open strings, we
anticipate the similarity to the `particle' case, since the
interaction with external gauge fields is restricted to the edges of
the world
sheet, which behave like tracks of moving particles.

We hope to look into the symmetry-breaking mechanism in closed-string
theory. We expect purely `stringy' effects in the closed-string model
which cannot be obtained in the models mentioned earlier. 
Indeed the subject of current
research concentrates on the closed-string theory, which is
phenomenologically interesting
\cite{8}; for a review see \cite{20}.  This is another reason for
investigating the mechanism in closed-string theory.

In this paper we will show a framework of
the symmetry-breaking mechanism in closed-string theory. The treatment
of the Hosotani mechanism is easy in the operator formalism \cite{1}.
We begin with the deformation of Virasoro generators and show the
interpretation of symmetry breaking in the model. The order
parameter is considered to be set by hand in this paper. The dynamical
determination of the order parameter is left to be
examined in future publications.

This paper is organised as follows. In \S 2 we review
the old (-fashioned) operator method \cite{1} and we describe torus
compactification in closed-string theory by this method. In
\S 3, firstly we introduce a gauge symmetry into closed-string
theory by current algebra in the usual way \cite{1}. Secondly, using an
analogy with the torus compactification, we formulate a
symmetry-breaking mechanism which can be interpreted as the Higgs
mechanism in the theory. The Hosotani mechanism in closed-string theory
is described in \S 4. We give an evaluation of a
four-point amplitude of scalars in the model. We also calculate the
same amplitudes in the Higgs model explained in \S 3. It is
shown that the difference can be seen in the amplitude even if the
masses of scalar bosons are identical in both models. The final section
is devoted to discussion.

In this paper we consider bosonic strings only. We set
$\alpha'=1/2$ throughout this paper.

\section{Torus compactification in closed-string theory}
We begin with a review of the operator approach to string
theory \cite{1}. First we introduce the Virasoro generators which are
constructed from the `harmonic oscillator' operators associated
with excitation of string modes:
\begin{equation}
L_n=\frac{1}{2}\sum_{m=-\infty}^\infty :\alpha_{n-m}\cdot\alpha_m:\,.
\end{equation}
Here the `normal order' should be taken as
\begin{eqnarray}
:\alpha_n^\mu\alpha_m^\nu:&=&\alpha_n^\mu\alpha_m^\nu
\qquad\mbox{when}\quad n<m\,\nonumber \\
&=&\alpha_m^\nu\alpha_n^\mu\qquad\mbox{when}\quad n>m\,.
\end{eqnarray}
By virtue of this treatment, $L_0$ is simply written as
\begin{equation}
L_0=\frac{1}{2}P^2+\sum_{m=1}^\infty
:\alpha_{-m}\cdot\alpha_m:\,
\end{equation}
where we rewrite $\alpha_0$ as $P$.

In closed string theory we must prepare a set of copies of
Virasoro generators $\tilde{L}_n$, which are expressed in right moving
oscillator modes $\tilde{\alpha}_m$.

Next we adopt a tachyonic state $|0\rangle$ as usual. It satisfies
\begin{equation}
L_0|0\rangle=\tilde{L}_0|0\rangle=|0\rangle\,.
\end{equation}

Now, since we wish to investigate gauge symmetry breaking in
subsequent sections, we ought to pay attention to the masses of
the light fields.

The `mass operator' is defined as
\begin{equation}
M^2=4(L_0+\tilde{L}_0-2)\,
\end{equation}
for closed strings. Thes quotation marks imply that this operator gives
masses for any external string state whose momentum is set equal to
zero.

In the closed strings, this mass operator is always
accompanied with the following mass-matching condition:
\begin{equation}
(L_0-\tilde{L}_0)|\cdot\rangle=0\,
\end{equation}
where $|\cdot\rangle$ is an external physical state. The physical
states are expected to be made from the `creation' operators
$\alpha_{-n}$ $(n>0)$ applied to the tachyonic state. To take a
complete account for
physical string states, it is 
necessary to mention the existence
of spurious states \cite{1}. However, we shall neglect that painful task
in this paper and merely remark that the construction of
spurious states is related to the Virasoro algebra.
  
Now consider compactification of one dimension to a circle,
$S^1$. We denote the dimension as the $I$th direction and the other
directions are labelled by the index $i$. The momenta of strings are
discretised in the $I$th direction. For closed strings it is well
known that there are windings on the torus; these are also specified by
an integer. These quantum numbers are rearranged according to momenta
of left- and right-moving sectors of strings;
\begin{equation}
P_L^I=\frac{\ell'}{2R}+mR\quad
\mbox{and}\quad
P_R^I=\frac{\ell'}{2R}-mR\qquad
\end{equation}
where $R$ is the radius of $S^1$. $\ell'$ and $m$ are integral
and represent discreteness of momentum on $S^1$ and the winding number
around 
$S^1$, respectively. The following two quantities will be important in
later analysis:
\begin{equation}
\frac{1}{2}\{(P_L^I)^2+(P_R^I)^2\}=\left(\frac{\ell'}{2R}\right)^2+
(mR)^2
\label{2.8}
\end{equation}
\begin{equation}
\frac{1}{2}\{(P_L^I)^2-(P_R^I)^2\}=\ell' m\,.
\label{2.9}
\end{equation}
Expression (\ref{2.8}) plus the contribution from a number of string
oscillators is just the mass of a certain external state.
Expression (\ref{2.9}) is closely connected with the mass-matching
condition on each string excitation level.

Let us consider a linear transformation in $P^I$. For
instance, suppose the following mixing:
\begin{equation}
\left(\begin{array}{c}
{P_L^I}'\\
{P_R^I}'\end{array}\right)=\left(\begin{array}{cc}
\cosh\alpha & \sinh\alpha\\
\sinh\alpha & \cosh\alpha\end{array}\right)\left(\begin{array}{c}
{P_L^I}\\
{P_R^I}\end{array}\right)
\label{2.10}
\end{equation}
where $\alpha$ is a constant. Then the quantities (\ref{2.8}) and
(\ref{2.9}) become
\begin{equation}
\frac{1}{2}\{({P^I_L}')^2+({P^I_R}')^2\}=\left(
\frac{\ell'}{2R}\right)^2 e^{2\alpha}+(mR)^2 e^{-2\alpha}
\label{2.11}
\end{equation}
and
\begin{equation}
\frac{1}{2}\{({P^I_L}')^2-({P^I_R}')^2\}=\ell'm
\label{2.12}
\end{equation}
under the transformation (\ref{2.10}).

As seen from (\ref{2.11}), this transformation gives rise to the
variation of the radius of $S^1$ according to
\begin{equation}
R\rightarrow R'=R\, e^{-\alpha}\,.
\end{equation}
Note that the combination (\ref{2.12}) (or (\ref{2.9})) is invariant
under mixing of this type. The exercise above can be regarded as a
very primitive example of a general lattice compactification
considered by many authors \cite{11}. The information of the momentum
lattice is usually assumed to be carried by external states. However,
there is  another possible formulation to describe this.

We consider a perturbation on the Virasoro operator. In the
case mentioned above, we define two operators $L_0'$ and $\bar{L}_0'$ as
follows:
\begin{eqnarray}
L_0'&=&L_0-\frac{1}{2}(P^I_L)^2+\frac{1}{2}({P^I_L}')^2\nonumber \\
\bar{L}_0'&=&\bar{L}_0-\frac{1}{2}(P^I_R)^2+\frac{1}{2}({P^I_R}')^2
\label{2.14}
\end{eqnarray}
where ${P^I}'$ is to be considered as an operator and is defined as
(\ref{2.10}). Thus the states do not have to carry all the information
about the compactification. In the present case, it is sufficient that
states have information of a fixed-radius compactification, say, $R =
1/\sqrt{2}$. The value of the radius $R$ is attributed to ${P^I}'$ in
${L_0}'$ of (\ref{2.14}).

The momentum in an extra space in string theory can be given
in two equivalent ways described above. Eventually, the replacement
$P_{L,R}^I\rightarrow {P_{L,R}^I}'$ in $L_n$
produces sets of Virasoro generators $L_n'$ and $\bar{L}_n'$ which obey
the same Virasoro algebra as
$L_n$.

The method described in this section will be extended to the
case with current-algebra symmetry in closed strings in the
subsequent sections.

\section{The Higgs mechanism}
First of all, we introduce current algebra into the closed
string theory \cite{1}. The current which has an index of the adjoint
representation can be analysed by the mode expansion on the
world-sheet coordinate:
\begin{equation}
J^a(z)=\sum_{-\infty}^\infty J^a_n\, z^{-z-1}\,.
\end{equation}
The coefficients, as the operators, satisfy the commutation
relations:
\begin{equation}
[J^a_m , J^b_n]=
if^{abc}J^c_{m+n}+\frac{1}{2}km\delta^{ab}\delta_{m+n,0}
\end{equation}
where the level of the Kac-Moody algebra $k$ is a constant. Gauge
symmetry arises from this (affine) Kac-Moody algebra. The zero
modes $J_0^a$ form a Lie algebra $G$ with the relation of the
generators
\begin{equation}
[J^a_0 , J^b_0]=if^{abc}J^c_{0}
\label{3.3}
\end{equation}
where $f^{abc}$ are the structure constants of $G$. We equip another
current $\bar{J}(\bar{z})$ for the right-moving modes. Therefore we can
describe the gauge symmetry $G\times\bar{G}$ in the low-energy sector of
the theory. In this paper we consider only the case with $\bar{G}\sim
G$. The massless spectrum consists of not only the graviton,
antisymmetric tensor field and dilaton but also non-Abelian
vector bosons and scalars. The string states
$J_{-1}^a\bar{\alpha}_{-1}^i|0\rangle$ and
$\alpha_{-1}^i\bar{J}_{-1}^a|0\rangle$ correspond to the gauge bosons
which transform as the (adj,1) and (1,adj) representation of
$G\times\bar{G}$. $J_{-1}^a\bar{J}_{-1}^b|0\rangle$ corresponds to
a multiplet of scalars of the (adj,adj) representations \cite{21,22}.
Here, and in the following, we choose $SU(2)$ as symmetry group $G$ for
simplicity. Moreover, we take the Cartan-Weyl basis, and then the
adjoint index $a=\{1,2,3\}$ is reformed to $\{+,-,3\}$. Namely,
(\ref{3.3}) yields $[J_0^+, J_0^-]=2J_0^3$, $[J_0^3, J_0^+]=J_0^+$,
and so on. The (unperturbed) Virasoro operator is expressed
as \cite{21,22,23,24}
\begin{equation}
L_0=\frac{1}{2}(P_L)^2+\frac{1}{2}\sum:\alpha_{-m}\cdot\alpha_m:
+\frac{1}{K}\sum:J_{-n}^aJ_n^a:\,,
\label{3.4}
\end{equation}
where $K=c_v+k$. $c_v$ is defined by the relation $f^{acd}f^{bcd}
=c_v\delta^{ab}$. $\bar{L}_0$ is written in an expression similar to
(\ref{3.4}).

Now we try to examine gauge symmetry breaking of this model
in analogy with the example illustrated in the preceding section.
We are going to use the following principles. First, we use the
Fourier components of original string modes and the Kac-Moody
currents as building blocks of new Virasoro operators ${L_0}'$ and
${\bar{L}_0}'$. Second, the perturbed Virasoro operators ${L_0}'$ and
${\bar{L}_0}'$ should involve continuous parameters which indicate the
deviation from the original ${L_0}$ and ${\bar{L}_0}$. Third, low-lying
external states, at least, are found trivially. Other points will be
will be clarified through the construction below.

The concept of perturbed operators has previously been considered
\cite{25}, but in a different context.

Similarly to the case with momenta on a circle, we consider the
following transformation among zero modes of currents belonging
to Cartan subalgebra:
\begin{eqnarray}
{J_0^3}{}'&=&\cosh\beta J_0^3+\sinh\beta\bar{J}_0^3\nonumber \\
{\bar{J}_0^3}{}'&=&\cosh\beta\bar{J}_0^3+\sinh\beta J_0^3
\end{eqnarray}
where, as previously
\[
J_0^a=\frac{1}{2\pi i}\oint dz\, J^a(z) 
\]
and $\beta$ is a parameter of perturbation.

Proceeding along this lines, we can construct new ${L_0}'$ and
${\bar{L}_0}'$:
\begin{eqnarray}
{L_0}'&=&L_0+A\{({J_0^3}')^2-({J_0^3})^2\}\nonumber \\
{\bar{L}_0}'&=&\bar{L}_0+A\{({\bar{J}_0^3}{}')^2-({\bar{J}_0^3})^2\}
\label{3.6}
\end{eqnarray}
where $A$ is a constant to be determined. This choice of the new
Virasoro operators makes the left-right level matching unchanged
(for any $A$),
\begin{equation}
{L_0}'-{\bar{L}_0}'={L_0}-{\bar{L}_0}\,.
\end{equation}
The constant $A$ is hence determined by consideration of other Virasoro
generators. They are naturally defined as
\begin{eqnarray}
{L_n}'&=&L_n+B\{{J_0^3}'{J_n^3}-{J_0^3}{J_n^3}\}\nonumber \\
{\bar{L}_n}'&=&\bar{L}_n+B\{{\bar{J}_0^3}{}'{\bar{J}_n^3}-
{\bar{J}_0^3}{\bar{J}_n^3}\}
\end{eqnarray}
where $B$ is a constant.

We assume that the new Virasoro generators (\ref{3.6}) obey the same
form of the Virasoro algebra
\begin{equation}
[{L_n}', {L_m}']=(n-m){L_{n+m}}'+\mbox{the central term}\,.
\label{3.9}
\end{equation}
Then the followings identities are required:
\begin{equation}
A=1/k\qquad B=2/k\,.
\label{3.10}
\end{equation}
Incidentally, the case with torus compactification is regained
formally by supposing that the gauge group is Abelian and $k=2$.

Now the mass-shifts for charged states are given by
\begin{equation}
\delta M_{op}^2=(4/k)
[2(\sinh\beta)^2\{(J_0^3)^2+(\bar{J}_0^3)^2\}+
4(\sinh\beta\cosh\beta)J_0^3\bar{J}_0^3]\,,
\label{3.11}
\end{equation}
where $J_0^3$ and $\bar{J}_0^3$ act as `charge' operators. Obviously,
gauge bosons ($J_{-1}^\pm\bar{\alpha}_{-1}^i|0\rangle$ and 
$\bar{J}_{-1}^\pm{\alpha}_{-1}^i|0\rangle$) and scalar bosons
($J_{-1}^\pm\bar{J}_{-1}^\pm|0\rangle$)
are found by making (\ref{3.11}) massive when $\beta$ is slightly
different from zero.

The physical meaning of this symmetry breaking can be read as
follows \cite{26}. The vertex operator for a scalar boson at zero
momentum is written in the form:
\begin{equation}
\phi^{ab}\sim\int dz d\bar{z}\, J^a(z) \bar{J}^b(\bar{z})\,.
\label{3.12}
\end{equation}
Therefore the second term proportional to $\sinh\beta\cosh\beta$ in
(\ref{3.11})
can be taken as the effect of condensation of zero-mode of $\phi^{33}$.
The `vacuum parameter' $\beta$ indicates the magnitude of the order
parameter $\langle\phi^{33}\rangle$ in some non-linear manner. In this
way, we can deal with a spontaneously broken ($SU(2)$) gauge theory with
massive gauge and Higgs bosons.

It is well known, on the other hand, that a level one ($k
=1$) $SU(2)\times SU(2)$ Kac-Moody symmetry can be generated from the
left- and right-moving modes of strings on a circle $S^1$ \cite{27}. To
see this, we only examine the left-moving algebra formed by the currents
defined as \cite{28}
\begin{eqnarray}
J^+(z)&=&:\exp(+i\sqrt{2}X_L^I):\nonumber \\
J^-(z)&=&:\exp(-i\sqrt{2}X_L^I):\nonumber \\
J^3(z)&=&(i/\sqrt{2})\partial_zX_L^I
\label{3.13}
\end{eqnarray}
where $X_L^I$ is the left-moving mode of strings on $S^1$. If the radius
of the circle $R$ is set to $1/\sqrt{2}$, they form $SU(2)$ Kac-Moody
algebra. If the value of $R$ deviates from $1/\sqrt{2}$, the $SU(2)$
symmetry will become broken and only $U(1)$ symmetry will be left,
which trivially exists in a model with a circle.

Alternatively, we can consider the background metric in path
integral as
\begin{equation}
G_{\mu\nu}=(-1,1,1,\ldots,R^2)
\end{equation}
and we take the string coordinate $X^I$ ($R=1/\sqrt{2}$). This results
in the same partition functions as before. Thus we can regard $R^2$ as
a zero mode of Kaluza-Klein scalar field; but in the string case,
this Kaluza-Klein scalar is interacting with the stringy
excitations which form $SU(2)$-adjoint scalar, and vector
and other massive fields. The deviation of $R$ from $1/\sqrt{2}$ can be
interpreted as the expectation value of a scalar field.

Thus one can check the mass spectrum in the `Higgs' mechanism
by comparison with each of the other models. The inclusion of background
fields is well described in the path integral approach \cite{29,30}. For
our purpose, we consider the zero-mode part of the partition
function in particular. The zero-mode piece of the bosonic
string coordinate on $S^1$ is of the form
\begin{equation}
\bar{X}^I=2\pi R (m\sigma_1+\ell\sigma_2)\,
\end{equation}
where $\sigma_1$ and $\sigma_2$ are coordinates of the world sheet and
$m$ and
$\ell$ are integers \cite{29}. Here we take the notion
$G_{\mu\nu}=(-1,1,\ldots,1)$ as in the former case above. Then the
integrand of the partition function is proportional to
\begin{equation}
\sum_{\ell m}e^{-S}=\left(\frac{\tau_2}{2R^2}\right)^{1/2}
\sum_{\ell'm}\exp\left(-\pi\tau_2\left(\frac{\ell'{}^2}{2R^2}+2m^2R^2\right)+2\pi
i\tau_1m\ell'\right)\,.
\label{3.16}
\end{equation}
We can read the mass spectrum from this expression, i.e.,
\begin{equation}
M^2=2(\ell'{}^2/2R^2+2m^2R^2-4)+\mbox{oscillators}
\label{3.17}
\end{equation}
where $-4$ in the parentheses comes from the tachyon in bosonic
strings. $\ell'$ and
$m$ are integers. The last term including $\tau_1$
in (\ref{3.16}) is concerned with the mass-matching constraint
such as (\ref{2.9}). In order to compare with our model, we define
\begin{equation}
Q_L\equiv(1/\sqrt{2})(\ell'+m)\qquad Q_R\equiv(1/\sqrt{2})(\ell'-m)
\end{equation}
and further, we set
\begin{equation}
Q'{}_L=\cosh\beta\,Q_L+\sinh\beta\,Q_R\qquad
Q'{}_R=\cosh\beta\,Q_R+\sinh\beta\,Q_L
\end{equation}
where $\sqrt{2}R=\exp(-\beta)$. Using this set of variables,
(\ref{3.17}) is rewritten as
\begin{equation}
M^2=4\{(Q'{}_L)^2+(Q'{}_R)^2-2\}+\mbox{oscillators}\,.
\label{3.20}
\end{equation}
Thus the mass shift is given by
\begin{equation}
\delta M^2= 4[(\sinh\beta)^2\{(Q_L)^2+(Q_R)^2\}+2(\sinh\beta\cosh
\beta)Q_LQ_R]\,.
\label{3.21}
\end{equation}
It seems that we may identify $Q_L$ and $Q_R$ with the eigenvalues of
charge operators $(2/k)^{1/2} J_0^3$ and $(2/k)^{1/2}\bar{J}_0^3$ with
$k=1$, respectively. Indeed, one can easily check the case for massless
external states. Namely, the gauge bosons
$J_{-1}^\pm\bar{\alpha}_{-1}^i|0\rangle$ have
$Q_L=\pm\sqrt{2}$ and $Q_R=0$, whilst
$\bar{J}_{-1}^\pm{\alpha}_{-1}^i|0\rangle$ have
$Q_L=0$ and $Q_R=\pm\sqrt{2}$. The scalar Higgs bosons
$J_{-1}^\pm\bar{J}_{-1}^\pm|0\rangle$ also acquire masses. We can
safely say that we can construct a modular invariant partition
function with Higgs mechanism in
$SU(2)\times SU(2)$ gauge theory, because of the correspondence with
torus compactification when
$k=1$.

\section{The Hosotani model in closed strings}
Let us now consider the Hosotani model in closed string
theory. The model we consider has $SU(2)\times SU(2)$ symmetry at first,
in the same fashion as discussed in \S 3, as well as a compact
extra-space $S^1$. First of all, we give a specific example for a
mixed transformation among internal momenta and charge operators.
That is,
\begin{equation}
\left(
\begin{array}{c}
P_L^I{}'\\
Q'{}_L\\
P_R^I{}'\\
Q'{}_R
\end{array}
\right)=
\left(
\begin{array}{cccc}
\cosh\theta & 0 & 0 & \sinh\theta\\
0 & \cosh\theta & \sinh\theta & 0\\
0 & \sinh\theta & \cosh\theta & 0\\
\cosh\theta & 0 & 0 & \cosh\theta
\end{array}
\right)
\left(
\begin{array}{c}
P_L^I\\
Q_L\\
P_R^I\\
Q_R
\end{array}
\right)
\label{4.1}
\end{equation}
where $Q_L=(2/k)^{1/2}J_0^3$ and $Q_R=(2/k)^{1/2}\bar{J}_0^3$, and the
notations are same as before, but the number of large dimensions is
one less than that in the Higgs model treated in \S 3.
Subsequently, we construct the Virasoro operators as follows:
\begin{eqnarray}
L'{}_0&=&L_0+\frac{1}{2}\{(Q'{}_L)^2+(P_L^I{}')^2-(Q_L)^2-(P_L^I)^2\}
\nonumber \\
\bar{L}'{}_0&=&\bar{L}_0+\frac{1}{2}
\{(Q'{}_R)^2+(P_R^I{}')^2-(Q_R)^2-(P_R^I)^2\}\,.
\end{eqnarray}
This example is clearly a straightforward extension
of the previous models in \S\S 2 and 3. The Virasoro generators
$L_n$ can also be constructed in a similar way.

Generally speaking, a transformation matrix $T$ which belongs
to $SO(2,2)$ can be utilised in this type of model. Namely,
if by such a matrix the relation
\begin{equation}
{}^tT\left(
\begin{array}{cccc}
1 & 0 & 0 & 0\\
0 & 1 & 0 & 0\\
0 & 0 & -1 & 0\\
0 & 0 & 0 & -1
\end{array}
\right)T
=
\left(
\begin{array}{cccc}
1 & 0 & 0 & 0\\
0 & 1 & 0 & 0\\
0 & 0 & -1 & 0\\
0 & 0 & 0 & -1
\end{array}
\right)
\end{equation}
is satisfied, then the condition $L'_0-\bar{L}'_0=L_0-\bar{L}_0$ holds
by use of
$T$ as the transformation matrix in (\ref{4.1}). As is known, the number
of parameters which characterise a general $T$ is six. However,
transformations which belong to a subgroup $SO(2)_L\times SO(2)_R$ for
instance, the rotation
\begin{equation}
\left(
\begin{array}{c}
P_L^I{}'\\
Q'{}_L\\
P_R^I{}'\\
Q'{}_R
\end{array}
\right)=
\left(
\begin{array}{cccc}
\cos\phi_L & -\sin\phi_L & 0 & 0\\
\sin\phi_L & \cos\phi_L & 0 & 0\\
0 & 0 & \cos\phi_R & -\sin\phi_R\\
0 & 0 & \sin\phi_R & \cos\phi_R
\end{array}
\right)
\left(
\begin{array}{c}
P_L^I\\
Q_L\\
P_R^I\\
Q_R
\end{array}
\right)
\label{4.4}
\end{equation}
leaves each Virasoro operator unchanged:
\begin{equation}
L'_0=L_0\qquad \bar{L}'_0=\bar{L}_0\,.
\label{4.5}
\end{equation}
In other words, the transformation such as (\ref{4.4}) makes no change
in the mass spectrum in the string model. As far as symmetry
breaking is concerned the whole mass spectrum is governed by four
parameters. Again, of these four parameters two correspond to the
mechanism considered before, in \S\S 2 and 3, one is concerned with the
variation of the size of
$S^1$ and one is concerned with the Higgs mechanism. Thus the number of
parameters which `describe' the geometrical symmetry breaking
is two. One of these two parameters is, say, $\theta$ in (\ref{4.1}).
The other will be discussed later, because we first wish to clarify
the physical implications of the example.

To see the physical interpretation, we think about the
difference in mass operators:
\begin{eqnarray}
(L'_0+\bar{L}'_0)-(L_0+\bar{L}_0)
&=&(\sinh\theta)^2\{(P_L^I)^2+(P_R^I)^2+(Q_L)^2+(Q_R)^2\}\nonumber \\
& &+2(\sinh\theta\cosh\theta)(P_L^IQ_R+Q_LP^I_R)\,.
\end{eqnarray}
In the limit of small $\theta$, this reduces to
\begin{equation}
(2\theta)(2/k)^{1/2}(P_L^I\bar{J}_0^3+P_R^I J_0^3)\equiv - V_0\,.
\end{equation}
This expression can be regarded as the `zero-mode' of the vertex,
of vector bosons:
\begin{equation}
(2/k)^{1/2}(\bar{J}(\bar{z})\partial X_L^I(z)+
J(z)\bar{\partial}X_R^I(\bar{z}))\, e^{ikX}\,.
\end{equation}
In addition, when we analyze the new propagator with $\theta\ne 0$, we
can give the following expansion in terms of the original propagator
($\theta=0$), up to the left-right matching constraint:
\begin{eqnarray}
\frac{1}{L'_0+\bar{L}'_0-2}
&=&\frac{1}{L_0+\bar{L}_0-2}+\frac{1}{L_0+\bar{L}_0-2}V_0
\frac{1}{L_0+\bar{L}_0-2}\nonumber
\\ &
&+\frac{1}{L_0+\bar{L}_0-2}V_0\frac{1}{L_0+\bar{L}_0-2}
V_0\frac{1}{L_0+\bar{L}_0-2}+\dots\,.
\end{eqnarray}
This shows the insertion of interactions with zero-modes of gauge
fields on $S^1$. Therefore we can say that the mechanism described
above is a simplest example of the Hosotani mechanism in closed-string
theory. It can be also said that in this example the scale of the
vacuum gauge field is given by
\begin{equation}
\langle A_I^3\rangle\sim(-2\theta)
\end{equation}
in the small $\theta$ limit. $A_I^3$ denotes a certain combination of
zero modes of two gauge fields.

Now let us examine scattering amplitudes in this
framework. A tree amplitude of four particles contains poles
associated with the intermediate states \cite{1}. Thus we can obtain
much knowledge about the mass spectrum and we can compare the
amplitude with that in other models such as the Higgs model. For
simplicity we show a scattering amplitude of four doubly adjoint
scalar bosons:
\begin{equation}
\phi^{+3}+\phi^{33}\rightarrow\phi^{+3}+\phi^{33}\,.
\end{equation}
The superscripts indicate the left and right $SU(2)$ `charge'
classified as in \S 3. The external states are represented as
\begin{equation}
\phi^{+3}\sim(2/k)J_{-1}^+\bar{J}_{-1}^3|0,P^I=0\rangle
\end{equation}
and the $\phi^{33}$ emission vertex is determined to be
\begin{equation}
V'{}^{33}(z=\bar{z}=1)=\frac{2}{k}\left(\sum_nJ_n^3-J_0^3+{J_0^3}'\right)
\left(\sum_n\bar{J}_n^3-\bar{J}_0^3+{\bar{J}_0^3}{}'\right)\, e^{ikX(1)}
\label{4.13}
\end{equation}
by investigation of the ghost-decoupling condition \cite{1}.

Using these preparations, we can calculate the four-point
amplitude of scalars and this reduces to:
\begin{eqnarray}
A(\theta)&=&\frac{\kappa^2}{4}\times\left[\left(
\frac{\Gamma\left(-\frac{1}{8}s+\frac{1}{8}m^2+1\right)
\Gamma\left(-\frac{1}{8}t-1\right)
\Gamma\left(-\frac{1}{8}u+\frac{1}{8}m^2+1\right)}{
\Gamma\left(\frac{1}{8}s-\frac{1}{8}m^2\right)
\Gamma\left(\frac{1}{8}t+2\right)
\Gamma\left(\frac{1}{8}u-\frac{1}{8}m^2\right)}
\right.\right.\nonumber
\\ 
& &+\frac{\Gamma\left(-\frac{1}{8}s+\frac{1}{8}m^2+1\right)
\Gamma\left(-\frac{1}{8}t-1\right)
\Gamma\left(-\frac{1}{8}u+\frac{1}{8}m^2-1\right)}{
\Gamma\left(\frac{1}{8}s-\frac{1}{8}m^2\right)
\Gamma\left(\frac{1}{8}t\right)
\Gamma\left(\frac{1}{8}u-\frac{1}{8}m^2\right)}\nonumber
\\
& &\left.+\frac{\Gamma\left(-\frac{1}{8}s+\frac{1}{8}m^2-1\right)
\Gamma\left(-\frac{1}{8}t-1\right)
\Gamma\left(-\frac{1}{8}u+\frac{1}{8}m^2+1\right)}{
\Gamma\left(\frac{1}{8}s-\frac{1}{8}m^2\right)
\Gamma\left(\frac{1}{8}t\right)
\Gamma\left(\frac{1}{8}u-\frac{1}{8}m^2\right)}\right)\nonumber
\\
&
&+\frac{2}{k}(\cosh\theta)^2\left(-
\frac{\Gamma\left(-\frac{1}{8}s+\frac{1}{8}m^2\right)
\Gamma\left(-\frac{1}{8}t-1\right)
\Gamma\left(-\frac{1}{8}u+\frac{1}{8}m^2\right)}{
\Gamma\left(\frac{1}{8}s-\frac{1}{8}m^2\right)
\Gamma\left(\frac{1}{8}t\right)
\Gamma\left(\frac{1}{8}u-\frac{1}{8}m^2\right)}\right.\nonumber
\\
& &+
\frac{\Gamma\left(-\frac{1}{8}s+\frac{1}{8}m^2\right)
\Gamma\left(-\frac{1}{8}t+1\right)
\Gamma\left(-\frac{1}{8}u+\frac{1}{8}m^2-1\right)}{
\Gamma\left(\frac{1}{8}s-\frac{1}{8}m^2\right)
\Gamma\left(\frac{1}{8}t\right)
\Gamma\left(\frac{1}{8}u-\frac{1}{8}m^2+1\right)}\nonumber
\\
&
&\left.\left.+
\frac{\Gamma\left(-\frac{1}{8}s+\frac{1}{8}m^2-1\right)
\Gamma\left(-\frac{1}{8}t+1\right)
\Gamma\left(-\frac{1}{8}u+\frac{1}{8}m^2\right)}{
\Gamma\left(\frac{1}{8}s-\frac{1}{8}m^2+1\right)
\Gamma\left(\frac{1}{8}t\right)
\Gamma\left(\frac{1}{8}u-\frac{1}{8}m^2\right)}\right)\right]
\label{4.14}
\end{eqnarray}
where $\kappa^2$ is the coupling in closed strings and
$m^2(\theta)=8/k(\sinh\theta)^2$. In the amplitude (\ref{4.14}), the
exchange of tachyon and graviton appears in the first term in the first
parentheses. This term has poles at $t=-8, 0, \dots$. The gauge
interaction and interaction among charged particles are contained in
the second parentheses. For instance, the last term describes the
contribution of massive gauge bosons to intermediate states. The
term contains poles in the $s$ channel at $s=m^2,m^2+8,\dots$. Note
that the ratio of couplings of gauge boson and graviton exchange
at tree level turns out to be
\begin{equation}
(2/k)(\cosh\theta)^2\,.
\end{equation}
The $k$ dependence has been mentioned by Ginsparg \cite{31}; a new
feature is the dependence of effective coupling on the `vacuum
parameter' $\theta$.

We can also carry out calculation of the same amplitude in
the Higgs model introduced in \S 3. The result is
\begin{eqnarray}
&
&\frac{4}{\kappa^2}A_{Higgs}(\beta)=\frac{4}{\kappa^2}A(\theta\rightarrow
\beta)\nonumber \\
& &+\frac{2}{k}(\sinh\beta)^2\left(-\frac{
\Gamma\left(-\frac{1}{8}s+\frac{1}{8}m^2\right)
\Gamma\left(-\frac{1}{8}t-1\right)
\Gamma\left(-\frac{1}{8}u+\frac{1}{8}m^2\right)}{
\Gamma\left(\frac{1}{8}s-\frac{1}{8}m^2\right)
\Gamma\left(\frac{1}{8}t\right)
\Gamma\left(\frac{1}{8}u-\frac{1}{8}m^2\right)}\right)\nonumber \\
& &+\left(\frac{2}{k}\right)^2(\sinh\beta)^2(\cosh\beta)^2\left(\frac{
\Gamma\left(-\frac{1}{8}s+\frac{1}{8}m^2\right)
\Gamma\left(-\frac{1}{8}t+1\right)
\Gamma\left(-\frac{1}{8}u+\frac{1}{8}m^2\right)}{
\Gamma\left(\frac{1}{8}s-\frac{1}{8}m^2+1\right)
\Gamma\left(\frac{1}{8}t\right)
\Gamma\left(\frac{1}{8}u-\frac{1}{8}m^2+1\right)}\right)
\label{4.16}
\end{eqnarray}
where $m^2=8/k(\sinh\beta)^2$. We find that the amplitudes
differ slightly between the two models, even though the masses of gauge
bosons are same. This difference is, of course, due to the ways of
coupling to the background fields, i.e., gauge and scalar bosons.
The last term in the additional terms in (\ref{4.16}) contains two
four-scalar couplings and two Higgs background fields
$\langle\phi^{33}\rangle$ in the scattering of low-lying states.

So far we have studied the Hosotani model using an example (\ref{4.1})
which includes only one vacuum parameter. We must return to
investigate another freedom in modification of mass spectrum.
Needless to say, two parameters originate from (the combinations
of) two vacuum gauge fields associated with $SU(2)_L$ and $SU(2)_R$
symmetry. Again, we give another example for finding the
parameter:
\begin{equation}
\left(
\begin{array}{c}
P_L^I{}'\\
Q'{}_L\\
P_R^I{}'\\
Q'{}_R
\end{array}
\right)=
\left(
\begin{array}{cccc}
1 & B/2 & 0 & -B/2\\
-B/2 & 1 & B/2 & 0\\
0 & B/2 & 1 & -B/2\\
-B/2 & 0 & B/2 & 1
\end{array}
\right)
\left(
\begin{array}{c}
P_L^I\\
Q_L\\
P_R^I\\
Q_R
\end{array}
\right)
\label{4.17}
\end{equation}where $B$ is the new vacuum parameter. Though the matrix
in (\ref{4.17}) looks a little bizarre, this rotation matrix
evidently obeys
$SO(2,2)$ symmetry. Also multiplication of the matrices of the same
form defines a small group. The investigation of the mass
operator with this parameter reveals a possibility that eight
massless vector bosons emerge when $B=1$. These bosons probably 
take the form of some mixture of $SU(3)$ gauge fields. This
enlargement of symmetry can be understood when $k=1$, using the
equivalent torus compactlflcation \cite{27,32}. In this case, we must
employ a two-torus as an internal space. Let us assume that the
background fields, that is to say, metric and antisymmetric
tensor fields on the torus, are
\begin{equation}
G_{mn}=
\left(
\begin{array}{cc}
\cosh 2\theta & -\sinh 2\theta \\
-\sinh 2\theta & \cosh 2\theta 
\end{array}
\right)
\label{4.18}
\end{equation}
and
\begin{equation}
B_{mn}=
\left(
\begin{array}{cc}
0 & B\\
-B & 0
\end{array}
\right)
\label{4.19}
\end{equation}
where $m$ and $n$ are the indices of the two-torus and $\theta$ and $B$
are constants. Note that $\det G_{mn}=1$. When we consider the partition
function of the compactified model, we can read the mass spectrum
from the path-integral form. The comparison will be made in a
 manner similar to that in \S 3, but now we have two compactified
dimensions; both radii of two scales are set to $R=1$. First we
can construct the eigenvalue of the `charge' from a set of
quantised numbers, of momentum and winding, in the direction of
{\it one} dimension. Next, we interpret that another dimension will
remain as a one-torus, that is, a circle. There the notation
of the left-right momenta is maintained. In consequence, we
obtain a model which can be compared with the Hosotani model in
the operator formulation.

In this situation mentioned above, one can calculate the
partition function by the path-integral method and after some slightly
tedious rearrangement using Jacobi's transformation, we find the
mass spectrum:
\begin{equation}
M^2=\frac{1}{2}\{({P_L^I}')^2+({P_R^I}')^2+(Q_L')^2+(Q_R')^2\}
+(\mbox{oscillators})\,.
\label{4.20}
\end{equation}
We have only to know how we can construct momenta $P'$ and charges
$Q'$ at finite $\theta$ and $B$ from these at $\theta=B=0$ (denoted as
$P$ and $Q$). The relation turns out to be:
\begin{eqnarray}
& &\left(
\begin{array}{c}
P_L^I{}'\\
Q'{}_L\\
P_R^I{}'\\
Q'{}_R
\end{array}
\right)=
\left(
\begin{array}{cccc}
\cosh\theta & 0 & 0 & \sinh\theta\\
0 & \cosh\theta & \sinh\theta & 0\\
0 & \sinh\theta & \cosh\theta & 0\\
\sinh\theta & 0 & 0 & \cosh\theta
\end{array}
\right)
\nonumber \\
& &\qquad\qquad\qquad\times
\left(
\begin{array}{cccc}
1 & B/2 & 0 & -B/2\\
-B/2 & 1 & B/2 & 0\\
0 & B/2 & 1 & -B/2\\
-B/2 & 0 & B/2 & 1
\end{array}
\right)
\left(
\begin{array}{c}
P_L^I\\
Q_L\\
P_R^I\\
Q_R
\end{array}
\right)\,.
\label{4.21}
\end{eqnarray}
The correspondence of parameters is apparent. The symmetry
enhancement (e.g., when $\theta=0$ and $B=1$) can be explained by
explicit construction of currents from bosonic coordinates
similar to (\ref{3.13}) in the compactified model. As is well known
the maximal symmetry group has rank equal to the dimension of
the torus. Note, however, that in our model in the operator
method, the maximal symmetry is a single $SU(3)$, not $SU(3)\times
SU(3)$; the latter is attained by the previous two-torus
compactification. There is not a one-to-one correspondence between the
spectrum of the strings with currents and that of the na\"{\i}ve
compactified string model. If we want exactly similar partition
functions, we need an appropriate projection in compactification of the
background. Anyway, the occurrence of symmetry enhancement is a
purely `stringy' effect.

This feature clearly originates from condensation of the
antisymmetric tensor field, as is manifestly shown in this
parametrisation; although this is less obvious when we take a
general method of parametrisation of the vacuum parameters.

\section{Discusson}
In this paper we have shown the method describing the gauge
symmetry breaking mechanism, especially the Hosotani mechanism, in
bosonic string theory by introduction of modified
Virasoro operators. Also, a comparison with the
models with symmetries which come from torus compactification has
been made.

We have left two problems which require further study. One is on the
derivation of Virasoro generators from the `first principles' such as
the consideration of stress tensor in two-dimensional theory.
The study of Wess-Zumino-Witten models will help the investigation of
some algebraic structure when background fields
exist. The progress in the study of the
case with $k>1$ will prove helpful in this problem.
It is necessary to research into the connection with the description of
conformal field theories.
Another problem, apart from the trivial extension to general symmetry
groups, is the application to the supersymmetric model.
We can examine the dynamical determination of
vacuum parameters and finite temperature effects based on such a model,
because the supersymmetric models do not suffer from
the disease of tachyon. In superstring models, we have two further
distinct points of interest. First, the enlargement of the gauge
group is anticipated in the models as in the bosonic case. In the 
supersymmetric case, however, projections onto physical states are
essential and then the symmetry enhancement mechanism may not have a
straightforward generalisation. Second, we expect
interesting aspects in the thermal property of the model. We
hope that the consideration of symmetry-breaking (or enhancement)
mechanisms in string theory brings a new perspective to string
cosmology.

Finally, we wonder how the Hosotani model model on a torus can be
extended to the Wilson loop mechanism on general non-simply-connected
manifolds. Perhaps an analogous construction of Wilson loops can
be performed in terms of the description of strings on
non-simply-connected Calabi-Yau manifolds by Gepner \cite{12}. In such a
model, if it is possible, vacuum parameters will no longer be continuous
quantities but discrete ones.

\section*{Acknowledgements}
The author thanks T. Itoh and A. Nakamula for discussions.

This work is supported in part by the Grant-in-Aid for Encouragement of
Young Scientists from the Ministry of Education, Science and Culture (\#
63790150).

The author is grateful to the Japan Society for the Promotion of
Science for a fellowship. He also thanks Iwanami F\=ujukai for
financial aid.


\end{document}